\documentclass[12pt]{article}

\usepackage{amsmath}
\usepackage{amssymb}
\usepackage{latexsym}
\usepackage{graphicx}
\usepackage{epsfig}
\usepackage{color}

\def\beq{\begin{equation}}
\def\eeq{\end{equation}}
\def\bea{\begin{eqnarray}}
\def\eea{\end{eqnarray}}

\begin{document}

\begin{titlepage}

\vspace*{1cm}
\begin{center}
{\bf \Large On the Localisation of 4-Dimensional\\[2mm] Brane-World Black Holes II:
the general case}

\bigskip \bigskip \medskip

{\bf P. Kanti}$^{1,2,\ddag}$, {\bf N. Pappas}$^2$ and {\bf T. Pappas}$^1$

\bigskip
$^1${\it Division of Theoretical Physics, Department of Physics,\\
University of Ioannina, Ioannina GR-45110, Greece}

$^2${\it Nuclear and Particle Physics Section, Physics Department,\\
University of Athens, Athens GR-15771, Greece}

\bigskip \medskip
{\bf Abstract}
\end{center}
We perform a comprehensive analysis of a number of scalar field theories in a
attempt to find analytically 5-dimensional, localised-on-the-brane, black-hole
solutions. Extending a previous analysis, we assume a generalised Vaidya ansatz
for the 5-dimensional metric tensor that allows for time-dependence, non-trivial
profile of the mass function in terms of the bulk coordinate and a deviation from the
over-restricting Schwarzschild-type solution on the
brane. In order to support such a solution, we study a variety of theories including
single or multiple scalar fields, with canonical or non-canonical kinetic terms,
minimally or non-minimally coupled to gravity. We demonstrate that for such a
metric ansatz and for a carefully chosen, non-isotropic in 5 dimensions, energy-momentum tensor,
solutions that have the form of a Schwarzschild-(Anti)de Sitter or Reissner-Nordstrom
type of solution do emerge, however, the resulting profile of the mass-function
along the bulk coordinate, when allowed, is not the correct one to eliminate the
bulk singularities.

\bigskip\bigskip\bigskip\bigskip\bigskip
\bigskip\bigskip\bigskip\bigskip\bigskip
\bigskip\bigskip\bigskip\bigskip\bigskip

$^{\ddag}$corresponding author: pkanti@cc.uoi.gr

\end{titlepage}

%%%%%%%%%%%%%%%%%%%%%%%%%%%%%%%%%%%%%%%%%%%%%%%%%%%%%%%%%%%%%%%%%%%%%%

\section{Introduction}
\label{Intro}

The intriguing proposal that our world may be a 4-dimensional hypersurface, a brane,
embedded in a higher-dimensional spacetime, the bulk, was put forward several decades
ago ~\cite{misha, akama} but was revived in the context of the more recent theories postulating
the existence of extra spacelike dimensions \cite{ADD, RS}. Although the initial objective was
to address the hierarchy problem, this was soon surpassed and these theories became the
framework in the context of which implications on gravity, particle physics and cosmology
were intensively investigated.

The question of how the extra spacelike dimensions affect the black-hole physics, in fact,
preceded the aforementioned brane-world theories, even the oldest ones. Black-hole solutions
in a spacetime with an arbitrary number of extra spacelike dimensions were derived and
their properties investigated already from the 1960's : higher-dimensional versions of the Schwarzschild
solution as well as of Kerr solutions were derived in \cite{Tangherlini, MP}, respectively. In these
studies, the extra dimensions were assumed to be flat or, at the most, uniformly curved due
to the presence of a bulk cosmological constant - in any case, all spacelike dimensions
played the same role in the structure of the gravitational background, a concept followed
also in the Large Extra Dimensions Scenario \cite{ADD}.

In the Warped Extra Dimensions Scenario \cite{RS}, on the other hand, the structure of spacetime
was distinctly different. The sole extra spacelike dimension could be finite or infinitely-long,
and our brane was a slice of a 5-dimensional Anti-de Sitter spacetime. More importantly,
the metric tensor along this particular fifth dimension exhibited a warping, a feature that
helped to either recast the hierarchy problem in the context of the two-brane model or
to localise gravity in the single-brane model.

How the properties of black holes emerging in the context of this theory are affected
by the particular structure of spacetime remains, to a large extent, unclear. The main
reason for this is that, despite the time and energy invested on this question (for reviews, see
\cite{reviews}), no analytical solution describing a regular, localised-on-the-brane,
5-dimensional black hole has been derived. The first such attempt appeared
in \cite{CHR} and the line-element had the form
%%%%%%%%%%%%%%%%%
\begin{equation}
ds^2=e^{2A(y)}\,\left[-\left(1-\frac{2M}{r}\right)dt^2 + \left(1-\frac{2M}{r}\right)^{-1} dr^2
+ r^2\,(d\theta^2+\sin^2\theta\,d\varphi^2)\right] + dy^2,
\label{black-string}
\end{equation}
%%%%%%%%%%%%%%
where $y$ denotes the coordinate along the fifth dimension and $A(y)$ is the warp factor.
In the above, $M$ was a constant quantity identified with the black-hole mass -- indeed, for
$A(y)=-k |y|$ \cite{RS}, with $k$ the AdS curvature, the projected line-element on the brane,
located at $y=0$, assumed the form of a Schwarzschild solution. However, the above
line-element does not describe a regular black hole from the 5-dimensional point-of-view:
the corresponding curvature invariant quantities are plagued by the presence of
singularities extending along the extra coordinate; for instance, we find that
%%%%%%%%%%%%
\bea
R^{MNRS}R_{MNRS} = \frac{48e^{-4A(y)}M^2}{r^6}+...\,.
\label{inv-RS}
\eea
%%%%%%%%%%%
The above reveals the existence of a singularity at $r=0$ at every point along the
extra dimension; in addition, for $A(y)=-k|y|$, or for any other decreasing warp function,
the above quantity increases in magnitude as we move away from the brane. In
the single-brane model, it actually becomes infinite at $y$-infinity -- this is indeed
ironic in the context of a model that aims at keeping gravity localised close to the
brane so that 4-dimensional physics is restored despite the presence of an infinitely
extended extra dimension. The above solution is clearly not a black-hole solution
but a black-string one, a class of solutions first discovered in the context of string
theory and found to be unstable both in the absence or presence of a bulk cosmological
constant \cite{GL, RuthGL}.

Since then, numerous attempts have been made in the literature to derive a regular,
5-dimensional black hole located on the brane (for some of them, see \cite{tidal, Papanto,
KT, KOT, CasadioNew, Frolov, Karasik, GGI, CGKM, Ovalle, Harko, daRocha}) but no such
analytical solution was derived in closed form. Although analytical solutions of this type were
found in the context of lower-dimensional models \cite{EHM, AS}, their 5-dimensional
analogues remained elusive. Numerical solutions did emerge \cite{KTN, Kudoh, TT}
in the context of five- and six-dimensional warped models but these described only black holes
with horizon radius smaller than or at most of the order of the AdS length $\ell =1/k$.
Several arguments that supported the idea of the existence \cite{Fitzpatrick, Zegers, Heydari, Dai}
or non-existence \cite{Bruni, Dadhich, Kofinas, Tanaka, EFK, EGK, Yoshino, Kleihaus}
of these black-hole solutions were largely expressed. Novel numerical solutions that appeared
in recent years \cite{FW, Page} describing both large and small black holes reinforced
the arguments in favour of the existence of brane-world black holes but the absence
of an analytic solution left the study of their properties incomplete.

In this work, we return to an idea that first appeared in \cite{KT,KOT} according to which
a regular, 5-dimensional, localised-on-the brane black hole may emerge if a new type
of line-element, one that would admit a non-trivial profile of the mass function along
the extra dimension, is assumed. In that case, if the constant quantity $M$
appearing in the line-element (\ref{black-string}) is replaced by a function $m=m(y)$
that decreases faster than $e^{4A}$, then the singularities present in the curvature
invariant quantities, as in Eq. (\ref{inv-RS}), would be eliminated hopefully at a short
distance from the brane. In \cite{KT,KOT}, it was also demonstrated that not any type
of brane line-element could be used; instead the Vaidya-type line-element was singled
out as the most appropriate one that keeps the number of bulk singularities at the
minimum. Such a line-element of course is not a vacuum solution: the non-trivial
dependence of the 4-dimensional part of the metric on the extra coordinate made the
introduction of some form of matter in the bulk imperative. Alternative attempts, either
analytical or numerical, to derive regular, localised black holes also assumed some form
of matter either in the bulk \cite{KT, KOT, Frolov, Dai} or on the brane \cite{GGI, CGKM,
Ovalle, Heydari, Kleihaus, Andrianov}, or even additional geometrical terms \cite{Papanto, Cuadros}.

The form of matter that was necessary to localise the black-hole topology close to the brane,
although shown to be well-behaved everywhere in the bulk, was left unspecified in the context
of \cite{KT,KOT} -- it was nevertheless demonstrated that
conventional forms of matter, expressed in the form of ordinary field theories involving
minimally-coupled scalar or gauge degrees of freedom could not perform this task.
In a recent work \cite{KPZ}, a variety of non-conventional scalar field theories were
investigated including single or multiple scalar fields either minimally or non-minimally
coupled to gravity. The background assumed the form of a generalised Vaidya-type of
line-element with a $y$-dependence as well as a time-dependence of the mass function --
the first would help to localise the black-string singularity while the second would
allow for potential non-static solutions to emerge.  Once again, it was found that even
these forms of matter could not support this type of background.

Given the existence of the localised black holes found numerically over the years and
the absence of any analytical expression that would describe them, here we return to
our previous analysis \cite{KPZ} and adopt the most general type of a Vaidya-type of
line-element: now, we allow the mass-function to have an additional dependence on
the radial coordinate. In this way, the projected-on-the-brane line-element is
allowed to deviate from the perhaps over-restricting Schwarzschild-type of background,
and assume one where additional terms, as the ones associated with an effective
cosmological constant or tidal charges, to be present. The time-dependence and
dependence on the fifth coordinate will be kept, as in our previous work, to make
our analysis as general as possible. Due to the radical change in the line-element,
we will study a plethora of theories starting from the simplest one, a model with
only a cosmological constant in the bulk; we will then move to a series of
theories with single or multiple minimally-coupled scalar fields, interacting or
non-interacting, with canonical or non-canonical kinetic terms; we will finish with
the theory of a non-minimally coupled scalar field theory coupled to the Ricci scalar
through a general coupling function. We will try to determine the features of the
theory that would allow for a viable solution to emerge, and we will investigate the type
of solutions, that we get in each case, their characteristics, and the possibility
to localise the black-string singularity close to the brane.

The outline of our paper is as follows: in section 2, we present the geometrical
background and discuss its main features. In Section 3, we address the simplest case
of a bulk filled only with a cosmological constant. In the following section, we proceed
to study a plethora of theories with minimally-coupled scalar fields, and in Section 5
we investigate the theory of a non-minimally coupled scalar field. We discuss our
results and present our conclusions in Section 6.

%%%%%%%%%%%%%%%%%%%%%%%%%%%%%%%%%%%%%%%%%%%%%%%%%%%%%%%%%%%%%%%%%%%%%

\section{The Geometrical Background}

Motivated by the attractive features of the Vaidya metric pointed out in a previous
work \cite{KOT}, here we will also assume that the 5-dimensional gravitational background
is given by the generalised form
\begin{equation}
ds^2=e^{2A(y)}\left[-\left(1-\frac{2m(v,r,y)}{r}\right)dv^2+2dvdr+
r^2(d\theta^2+sin^2\theta\,d\phi^2)\right]+dy^2\,.
\label{vaidya-metric}
\end{equation}
In the above, apart from the $(v,y)$-dependence introduced in \cite{KOT}, an additional
dependence on the radial coordinate is postulated that will allow our metric background
to deviate from the Schwarzschild form. As mentioned in the Introduction, such a modification
may allow for terms proportional to an effective cosmological constant or to terms of
various forms associated with tidal charges to emerge -- terms of this type arise often in the
context of 5-dimensional or 4-dimensional effective (projected-on-the-brane) analyses;
demanding that these terms are zero, as was implicitly assumed in our previous work
\cite{KPZ}, may have caused the over-restriction of the set of field equations and,
unavoidably, the elimination of a viable solution.

The components of the Einstein tensor $G^M_N$ for the aforementioned generalised
form of the Vaidya metric (\ref{vaidya-metric}) are:
%
%%%%%%%%%%%%
\bea
&& G^{v}_{v} = G^{r}_{r} = 6A'^2 + 3A^{''}-\frac{2}{r^2}\,e^{-2A}\,\partial_{r}m\,,
\nonumber \\[1mm]
&& G^{\theta}_{\theta} = G^{\phi}_{\phi} = 6A'^2 + 3A^{''}
-\frac{1}{r}\,e^{-2A}\,\partial^2_{r}m\,, \nonumber \\[1mm]
&& G^{r}_{\,\,v} = \frac{2}{r^2}\,e^{-2A} \partial_{v}m -
\frac{1}{r}\,( \partial^2_{y}m + 4A'\partial_{y}m )\,, \label{einstein-tensor} \\[1mm]
&& G^{y}_{\,\,v} = e^{2A}\,G^{r}_{\,\,y}=\frac{1}{r^2}\,\partial_{y}m +
\frac{1}{r}\,\partial_{y}\partial_{r}m\,, \nonumber \\[1mm]
&& G^{y}_{y} = 6A'^2 - e^{-2A} \bigl(\frac{2}{r^2}\,\partial_{r}m +
\frac{1}{r}\,\partial^2_{r}m \bigr)\,,\nonumber \\[3mm]
&& G^v_{\,\,r}=G^y_{\,\,r}=G^v_{\,y}=0\,. \nonumber
\eea
%%%%%%%%%%%
We observe that as expected, for $\partial_{r}m=0$, the above expressions smoothly reduce
to the ones of the $r$-independent case studied in \cite{KPZ}.

The expressions of the 5-dimensional curvature invariant quantities will also play a
central role in our analysis: any viable solution should eliminate any singularities in the
bulk at a  finite distance from the brane thus leaving behind a regular Anti-de Sitter
spacetime and a localised 5-dimensional black hole solution close to the brane. For
our metric ansatz (\ref{vaidya-metric}), these are given by the following expressions
%%%%%%%%%%%%%
\bea R = -20A'^{2}-8A^{''}+ \frac{2e^{-2A}}{r}\bigl(\partial^2_{r}m +
\frac{2}{r}\,\partial_{r}m\bigr)
\label{Ricci-scalar},
\eea
%%%%%%%%%%%%%
\bea
R_{MN}R^{MN} &=& 80A'^{4} + 64A'^{2}A^{''} + 20A^{''2}  \nonumber\\[2mm]
&-&  \frac{4e^{-2A}}{r} \bigl(\partial^2_{r}m +
\frac{2}{r}\,\partial_{r}m\bigr)(A^{''}+4A'^{2}) +
\frac{2e^{-4A}}{r^2} \bigl[(\partial^2_{r}m)^2 +
\frac{4}{r^2}(\partial_{r}m)^2\bigr],
\label{Ricci-curvature}
\eea
%%%%%%%%%%%%%
and finally
%%%%%%%%%%%%%
\bea
R_{MNRS}R^{MNRS}
=  40A'^{4} + 32A'^{2}A^{''} + 16A^{''2} +\frac{48e^{-4A}m^2}{r^6}
-\frac{8A'^2e^{-2A}}{r}\bigl(\partial^2_{r}m + \frac{2}{r}\,\partial_{r}m\bigr)
\nonumber\\[2mm]
+ \frac{4e^{-4A}}{r^2}\Bigl[(\partial^2_{r}m)^2 +\frac{4m}{r^2}(\partial^2_{r}m-
\frac{4}{r}\,\partial_{r}m) -\frac{4}{r}\,\partial_{r}m\,\partial^2_{r}m+
\frac{8}{r^2}\,(\partial_{r}m)^2\Bigr]\,.
\label{Riemann-curvature}
\eea
%%%%%%%%%%%%%
For a constant mass function $m$, the above expressions reduce to the ones for
a black string \cite{CHR} with an infinitely extended singularity along the fifth
dimension. For $m=m(y)$ only, the first two invariants are everywhere well-defined
while the third one still has a singular term of the form $e^{-4A}m^2/r^6$;
an appropriately chosen function $m(y)$, i.e. decreasing faster than $e^{4A}$
with $y$, could help localise the black-hole singularity close to the brane,
unfortunately, the set of field equations were shown not to accept such a
solution in the context of a variety of field theory models \cite{KT, KOT, KPZ}.
As mentioned above, here, we will extend our assumption so that our metric
function admits forms that may deviate from the Schwarzschild form;
although this will make the system of field equations more flexible, the
$r$-dependence of the mass function leads to the appearance of additional
singular terms in the curvature invariant quantities. However, these terms
could also be eliminated, similarly to the black-string singular terms, if a
solution with a fast-enough decreasing profile of the mass function is found.

In the following sections, we consider a variety of scalar field theory models
and investigate whether these theories admit solutions whose geometrical
background is described by the line-element (\ref{vaidya-metric}).

%%%%%%%%%%%%%%%%%%%%%%%%%%%%%%%%%%%%%%%%%%%%%%%%%%%%%%%%%%%%

%%%%%%%%%%%%%%%%%%%%%%%%%%%%%%%%%%%%%%%%%%%%%%%%%%%%%%%%%%%%%%%%

\section{A Bulk filled with a Cosmological Constant} \label{constant}

For the sake of a comprehensive analysis, we start our study with the case of a bulk
that contains no scalar fields but only a cosmological constant $\Lambda_B$. As we
will see, this model cannot support a black-hole spacetime of the form (\ref{vaidya-metric}).
However, this case and some of its constraints will serve as prototypes from which the
subsequently considered, more elaborate, models will try to deviate in an effort to
find a viable solution.

The action functional in this case will simply be
%%%%%%%%%%
\bea
\mathcal{S} = \int d^4x\,dy\,\sqrt{-g} \,\left(\frac{R}{2 \kappa_5^2}-\mathcal{L}_{m}\right),
\label{action-basic}
\eea
%%%%%%%%%%
where, $g_{MN}$ and $R$ are the metric tensor and Ricci scalar, respectively, of the
5-dimen\-sional spacetime (\ref{vaidya-metric}), and $\kappa_5^2=8\pi G_N$ the
5-dimensional gravitational constant. Also, in this case, the general Lagrangian associated
with the distribution of matter/energy in the spacetime is $\mathcal{L}_{m}=\Lambda_B$,
where $\Lambda_B$ the bulk cosmological constant. The field equations resulting from
the aforementioned action have the form
%%%%%%%%%
\bea
G_{MN} \equiv R_{MN}-\frac{1}{2}\,g_{MN}\,R =
\kappa^2_{5}\,T_{MN}\,, \label{eq-motion}
\eea
%%%%%%%%%
with $T_{MN}$ being the energy-momentum tensor defined as
%%%%%%%%
\beq
T_{MN} \equiv \frac{2}{\sqrt{-g}}\,\frac{\delta (\sqrt{-g}\,\mathcal{L}_{m})}
{\delta g^{MN}} =-g_{MN}\,\Lambda_B\,.
\eeq
%%%%%%%%%

The cosmological constant introduces the same contribution to all diagonal, mixed components
of $T_{MN}$: $T^v_v=T^r_r=T^\theta_\theta=T^\varphi_\varphi=T^y_y=-\Lambda_B$.
A similar relation should then hold for the corresponding components of the Einstein
tensor; then, from the constraint $G^v_v=G^\theta_\theta$ and the expressions
(\ref{einstein-tensor}), we obtain the simple equation
%%%%%%%%%%%%%%%%%%%%%
\beq
\partial_r^2 m=\frac{2}{r}\,\partial_r m\, \label{mass-eq1}
\eeq
%%%%%%%%%%%%%%%%%%%%%%
with solution
%%%%%%%%%%%%%%
\beq
m(v,r,y)=B(v,y)\,r^3 + C(v,y)\,. \label{mass-sol1}
\eeq
%%%%%%%%%%%%%
Therefore, the metric tensor in (\ref{vaidya-metric}), describes a solution with a $C(v,y)/r$
Schwarzschild-type term, with a non-trivial profile along the $y$-coordinate, and an
additional term of ${\cal O}(r^2)$ typical of a cosmological constant. Although this
Schwarzschild-(Anti)de Sitter solution is a physically interesting one, describing a black
hole on the brane and a warped black string off the brane, it is not supported by the
off-diagonal components: the equation $G^y_{\,\,v}=T^y_{\,\,v}$, or, equivalently
%%%%%%%%%%%%%%%
\beq
\frac{1}{r^2}\,\partial_{y}m + \frac{1}{r}\,\partial_{y}\partial_{r}m=0\,,
\label{G_yv-constant}
\eeq
%%%%%%%%%%%%%%%
leads to the result $4r^3 \partial_y B + \partial_y C=0$, which cannot be satisfied
unless $\partial_y B=\partial_y C=0$. However, this means that the non-trivial
profile of the mass function along the extra coordinate necessary to localise the
black-hole singularity close to the brane is eliminated. The above result holds for
any sign or value of the cosmological constant $\Lambda_B$, and thus it is valid
for all cases of 5-dimensional (Anti)-de Sitter or Minkowski spacetimes.

%%%%%%%%%%%%%%%%%%%%%%%%%%%%%%%%%%%%%%%%%%%%%%%%%%%%%%%%%%%%%%%%

\section{A Field Theory with minimally-coupled Scalars} \label{minimal-gen}

In this section, we will study a variety of models with minimally-coupled scalar
fields with a general form of Lagrangian. We will start with the case of a single
scalar field and then proceed to the case of two interacting fields -- in both
cases, a general kinetic term will be considered, that will allow for both canonical and
non-canonical kinetic terms, as well as a general potential.

%%%%%%%%%%%%%%%%%%%%%%%%%%%%%%%%%%%%%

\subsection{A single scalar field with a general Lagrangian}

In the case of a single scalar field $\phi$, the general Lagrangian ${\cal L}_m$ of
the action (\ref{action-basic}) will now be replaced by
%%%%%%%%%%%%%%%
\begin{equation}
\mathcal{L}_{\phi}=\sum_{n=1}\,f_n(\phi)\left(\partial^M\phi\,\partial_M\phi\right)^n+V(\phi)\,,
\label{single-scalar}
\end{equation}
%%%%%%%%%%%%%%%%
where $f_n(\phi)$ are arbitrary, smooth functions of the scalar field $\phi$, and
$V(\phi)$ a general potential that may include the cosmological constant $\Lambda_B$.
For $n=1$ and $f_1(\phi)=1$, the above reduces to a Lagrangian of a scalar field with
a canonical kinetic term; for arbitrary $n$, it describes a general field theory of a
scalar field with a mixture of canonical and non-canonical kinetic terms. The
analysis, aiming to investigate whether such a theory supports a spacetime of
the form (\ref{vaidya-metric}), follows the same lines independently of the form
of the kinetic term -- therefore, here, we keep the most general expression given
by Eq. (\ref{single-scalar}) and present a unified analysis.

The Lagrangian (\ref{single-scalar}) leads to the following expression for the
energy-momentum tensor
%%%%%%%%%%%%%%%%
\begin{equation}
T_{MN}=2\sum_{n=1}n\,f_n(\phi)\left(\partial^P\phi\,\partial_P\phi\right)^{n-1}\partial_M\phi\,\partial_N\phi-
g_{MN}  \,\mathcal{L}_{\phi}.
\end{equation}
%%%%%%%%%%%%%%%%
The scalar field is assumed to be spherically symmetric and, in principle, to depend on
all remaining coordinates, i.e. $\phi=\phi(v,r,y)$. However, an important simplification
is imposed through the off-diagonal equation $G^v_{\,\,r}=T^v_{\,\, r}$ that takes the
explicit form
%%%%%%%%%%%%%%%%%%%%%%%
\begin{eqnarray}
2\,e^{-2A}\,\sum_{n=1}n\,f_n(\phi)\left(\partial^P\phi\,\partial_P\phi\right)^{n-1}(\partial_r\phi)^2=0\,.
\end{eqnarray}
%%%%%%%%%%%%%%%%%%%%
For a non-trivial warp factor and a non-trivial scalar kinetic term, the above demands that
the scalar field be not a function of the radial coordinate, i.e. $\partial_r\phi=0$.
Assuming henceforth that $\phi=\phi(v,y)$ and turning to the diagonal components
$G^v_v=T^v_v$ and $G^\theta_\theta=T^\theta_\theta$, we obtain\footnote{For simplicity,
we will henceforth set $\kappa^2_5=1$.}
%%%%%%%%%%%%%%%%%%
\bea
&& 6A'^2 + 3A^{''}-\frac{2}{r^2}\,e^{-2A}\,\partial_{r}m = -{\cal L}_\phi\,,\\[1mm]
&& 6A'^2 + 3A^{''}-\frac{1}{r}\,e^{-2A}\,\partial^2_{r}m=-{\cal L}_\phi\,,
\eea
%%%%%%%%%%%%%%%%
respectively. Combining the above two equations, we arrive again at the differential
equation (\ref{mass-eq1}) for the mass function and thus to the general solution
(\ref{mass-sol1}). Using this solution into the off-diagonal equation
$G^y_{\ v}=T^y_{\ v}$, we find
%%%%%%%%%%%%%%%
\beq
4r \partial_yB + \frac{\partial_y C}{r^2}=
2\sum_n n\,f_n(\phi)\,(\partial_P\phi\,\partial^P\phi)^{n-1}\,\partial_y\phi\,\partial_v\phi\,.
\end{equation}
%%%%%%%%%%%%%%%%%
Since the scalar field is independent of $r$, the right-hand-side of the above equation
will be also $r$-independent. But, due to the explicit dependence of the left-hand-side on
the radial coordinate, a mathematical inconsistency immediately arises. The only way
forward would be to assume that the mass functions $B$ and $C$ are again
$y$-independent which, however, is in contradiction to our basic assumption for
the behaviour of the metric function.

We notice that, despite our general ansatz for the Lagrangian describing the dynamics
of the scalar field -- note that the form of the potential was also kept general,
this model shared a basic characteristic with the (A)dS-Minkowski case considered
in section \ref{constant}: due to the constraint $\partial_r \phi=0$, it also satisfied the
property $T^v_v=T^\theta_\theta$; this inevitably led again to Eq. (\ref{mass-eq1}) and to
the solution (\ref{mass-sol1}). In an attempt to find a viable model, in the next
subsection, we present a two-scalar field theory that deviates from the aforementioned,
restrictive property of the energy-momentum tensor.

%%%%%%%%%%%%%%%%%%%%%%%%%%%%%%%%%%%%%%%%%%%%%%%%%%%%%%%%%%%%%%%%%%%%%%%%%%%%%%%%%%%%%%%%%

\subsection{Two interacting scalar fields}

We will first investigate the case of two interacting scalar fields with canonical kinetic
terms -- we will extend our analysis to cover the case of non-canonical kinetic
terms at the end of this subsection.  The dynamics and interactions of the two fields
$\phi$ and $\chi$ are described by the following Lagrangian
%%%%%%%%%%%%%%%%%%%
\begin{equation}
\label{action_two_1}
\mathcal{L}_{sc}=f^{(1)}(\phi,\chi)\,\partial^M\phi\,\partial_M\phi+
f^{(2)}(\phi,\chi)\,\partial^M\chi\,\partial_M\chi+V(\phi,\chi)\,,
\end{equation}
%%%%%%%%%%%%%%%%%
where $f^{\,(1,2)}$ are arbitrary smooth functions of the two fields, and
$V(\phi,\chi)$ a general potential. The corresponding energy-momentum tensor
has the form
%%%%%%%%%%%%%%%%
\begin{equation}
T_{MN}=2f^{(1)}(\phi,\chi)\,\partial_M\phi\,\partial_N\phi+
2f^{(2)}(\phi,\chi)\,\partial_M\chi\,\partial_N\chi-g_{MN}\,\mathcal{L}_{sc}\,.
\label{Tmn-2scalars}
\end{equation}
%%%%%%%%%%%%%%%%

The vanishing of the off-diagonal components $T^v_{\ r}$, $T^y_{\ r}$ and $T^v_{\ y}$,
due to the vanishing of the corresponding components of the Einstein tensor
(\ref{einstein-tensor}), now leads to two independent constraints, namely
%%%%%%%%%%%%%
\begin{eqnarray}
\label{constr_1}
&&f^{(1)}(\phi,\chi)\,(\partial_r\phi)^2+f^{(2)}(\phi,\chi)\,(\partial_r\chi)^2=0\,, \\[2mm]
\label{constr_2}
&&f^{(1)}(\phi,\chi)\,\partial_r\phi\partial_y\phi+f^{(2)}(\phi,\chi)\,\partial_r\chi\partial_y\chi=0\,.
\end{eqnarray}
%%%%%%%%%%%%%%%%
On the other hand, the diagonal field equations $G^v_v=T^v_v$ and
$G^\theta_\theta=T^\theta_\theta$ now take the form
%%%%%%%%%%%%%%%%%%
\bea
&& \hspace*{-1cm} 6A'^2 + 3A^{''}-\frac{2}{r^2}\,e^{-2A}\,\partial_{r}m =
2 e^{-2A}\,\Bigl[f^{(1)}(\phi,\chi)\,\partial_r\phi\,\partial_v \phi+
f^{(2)}(\phi,\chi)\, \partial_r \chi \partial_v \chi\Bigr]-{\cal L}_{sc}\,,
\label{vv-two}\\[1mm]
&& \hspace*{-1cm} 6A'^2 + 3A^{''}-\frac{1}{r}\,e^{-2A}\,\partial^2_{r}m=-{\cal L}_{sc}\,,
\label{Gthth-two}
\eea
%%%%%%%%%%%%%%%%
respectively. We observe that no constraint demands the vanishing of the expression
inside the square brackets on the right-hand-side of Eq. (\ref{vv-two}) and thus the
restrictive condition $T^v_v=T^\theta_\theta$ is now avoided. Rearranging the above
two equations, we obtain
%%%%%%%%%%%%%
\beq
 \frac{1}{r}\,\partial^2_{r}m - \frac{2}{r^2}\,\partial_{r}m =
2 \Bigl[f^{(1)}(\phi,\chi)\,\partial_r\phi\,\partial_v \phi+
f^{(2)}(\phi,\chi)\, \partial_r \chi \partial_v \chi\Bigr]\,. \label{mass-eq2}
\eeq
%%%%%%%%%%%%%

It is straightforward to exclude the following particular cases:

\begin{itemize}
\item{} one of the two fields does not depend on the radial coordinate: if, for
example, $\partial_r \phi=0$, then necessarily, from the constraint (\ref{constr_1}),
$\partial_r \chi=0$, too. But in that case, Eq. (\ref{mass-eq2}) reduces to (\ref{mass-eq1})
leading again to the solution (\ref{mass-sol1}). For the present model, the
$(^y_{\ v})$ component of the field equations has the form
%%%%%%%%%%%%%
\begin{equation}
\frac{\partial_y m}{r^2} + \frac{\partial_r \partial_y m}{r}=2 \left[f_1^{(1)}(\phi,\chi)\partial_y\phi\partial_v\phi+f_1^{(2)}(\phi,\chi)\partial_y\chi\partial_v\chi\right],
\label{G_yv}
\end{equation}
%%%%%%%%%%%%%
with the right-hand-side being again $r$-independent. However, using the
solution (\ref{mass-sol1}) on the left-hand-side and demanding mathematical
consistency, we are led again to the conditions $\partial_y B=\partial_y C=0$,
that unfortunately eliminate the assumed $y$-dependence of the mass function.

\item{} one of the two fields does not depend on the extra coordinate: if, for
example, $\partial_y \phi=0$, then, from the constraint (\ref{constr_2}), we
obtain either $\partial_r \chi=0$ or $\partial_y \chi=0$. According to the analysis
above, the first choice leads to a mathematical inconsistency, therefore we should
choose $\partial_y \chi=0$. In that case, the right-hand-side of Eq. (\ref{G_yv})
vanishes and, upon integration, the following solution is obtained
%%%%%%%%%%%%%
\beq
m(v,r,y)=\frac{E(v,y)}{r}+D(v,r)\,.
\eeq
%%%%%%%%%%%%%
The metric tensor then includes a modified Schwarzschild term $D(v,r)/r$ together
with a Reissner-Nordstrom-type term $E(v,y)/r^2$. Employing the above
expression for the mass function in Eq. (\ref{mass-eq2}), and demanding that
the right-hand-side is not a function of the $y$-coordinate, we are inevitably led
to the condition that the function $E$, and therefore the mass function altogether,
should also be $y$-independent.

\end{itemize}

From the analysis above, it is clear that if one of the two fields were not to depend
on either $r$ or $y$, neither would the other one. This is not the case with the
dependence on $v$ -- no inconsistency arises if only one of the fields is assumed
to be $v$-dependent. Nevertheless, one of the fields must necessarily do so,
otherwise Eqs. (\ref{mass-eq2}) and (\ref{G_yv}) reduce to (\ref{mass-eq1}) and
(\ref{G_yv-constant}), respectively, of Section 3 leading to a non-viable solution.
Similarly to the simpler case considered in \cite{KPZ}, it seems that if a solution
exists with a non-trivial profile for the mass-function along the extra coordinate,
the corresponding scalar-field configuration must necessarily be dynamical.

In what follows, and in order to address the most general case, we will assume
that the two fields depend on all coordinates, i.e. $\phi=\phi(v,r,y)$ and $\chi=\chi(v,r,y)$.
Focusing on the two off-diagonal constraints (\ref{constr_1})-(\ref{constr_2}),
we observe that, by solving the first one in terms of one of the coupling functions
$f^{\,(1,2)}$, namely
%%%%%%%%%%%%
\beq
f^{(1)}=-f^{(2)}\,\frac{(\partial_r \chi)^2}{(\partial_r \phi)^2}\,,
\label{f12}
\eeq
%%%%%%%%%%%
and substituting into the second, we obtain the alternative constraint
%%%%%%%%%%%%%%%
\beq
\partial_r\phi \,\partial_y\chi-\partial_r\chi\, \partial_y\phi=0\,. \label{constr_3}
\end{equation}
%%%%%%%%%%%%%%%
There is finally an additional diagonal component of the field equations, namely
$G^y_y=T^y_y$, that we have not considered yet. This takes the explicit form
%%%%%%%%%%%%%%%
\beq
\hspace*{-1cm} 6A'^2 -\frac{e^{-2A}}{r}\Bigl(\partial_r^2 m +
\frac{2}{r}\,\partial_{r}m\Bigr)=
2 \Bigl[f^{(1)}(\phi,\chi)\,(\partial_y\phi)^2+
f^{(2)}(\phi,\chi)\,(\partial_y \chi)^2\Bigr]-{\cal L}_{sc}\,.
\label{Gyy-two}
\eeq
%%%%%%%%%%%%%%%%%
However, using the relations (\ref{f12}) and (\ref{constr_3}) in the above equation,
we find that the expression inside the square brackets on the right-hand-side trivially
vanishes. Then, after combining Eqs. (\ref{Gthth-two}) and (\ref{Gyy-two}), we obtain
the simple equation
%%%%%%%%%%%%%
\beq
\frac{2}{r^2}\,\partial_r m=-3A''\,e^{2A}\,,
\label{mass-eq3}
\eeq
%%%%%%%%%%%%%
that, upon integration with respect to $r$, yields
%%%%%%%%%%%%%
\beq
m(v,r,y)=-\frac{A''}{2}\,e^{2A} r^3 +m_0 (v,y)\,.
\label{mass-sol3}
\eeq
%%%%%%%%%%%%
We observe that we obtain again the Schwarzschild-(A)dS type of solution for the
metric function, however, once again this turns out to be incompatible with the
remaining equations. In particular, when the above is substituted into Eq. (\ref{mass-eq2}),
the left-hand-side trivially vanishes leading to the additional constraint
%%%%%%%%%%%%%%%
\beq
f^{(1)}\,\partial_r\phi \,\partial_v\phi+ f^{(2)}\,\partial_r\chi\,\partial_v\chi=0\,, \label{constr_4}
\end{equation}
%%%%%%%%%%%%%%%
or, by using Eq. (\ref{f12}),
%%%%%%%%%%%%%%%
\beq
\partial_r\phi \,\partial_v\chi-\partial_r\chi\, \partial_v\phi=0\,. \label{constr_5}
\end{equation}
%%%%%%%%%%%%%%%
When, Eqs. (\ref{constr_3}) and (\ref{constr_5}) are used in Eq. (\ref{G_yv}), its
right-hand-side vanishes, too. Employing then the solution for the mass function
(\ref{mass-sol3}) on its  left-hand-side, we obtain the condition
%%%%%%%%%%%%%
\beq
-2 r \partial_y(A'' e^{2A}) + \frac{1}{r^2}\,\partial_y m_0=0\,,
\eeq
%%%%%%%%%%%%
which is again inconsistent with the assumptions of the model. Therefore, the theory
of two interacting scalar fields with canonical kinetic terms fails to support a viable
solution. Note that, although this particular field theory was constructed so that, in
principle, it avoids the restrictive condition $T^v_v=T^\theta_\theta$, the system
of field equations itself imposed this condition, or equivalently the constraint
(\ref{constr_4}).

We may quite easily demonstrate that the above negative result holds also in the case
where non-canonical kinetic terms are assumed for the two scalar fields. The Lagrangian
of the theory then reads
%%%%%%%%%%%%%%%
\begin{equation}
{\cal L}_{sc}= \sum_{n=1}f_n^{(1)}(\phi,\chi)\left(\partial^M\phi\partial_M\phi\right)^n+
\sum_{n=1}f_n^{(2)}(\phi,\chi)\left(\partial^M\chi\partial_M\chi\right)^n+V(\phi,\chi)\,,
\label{Lagr_2non}
\end{equation}
while the energy momentum tensor assumes the form
\begin{eqnarray}
T_{MN}&=&2\sum_{n=1}n\,f^{(1)}_n(\phi,\chi)\left(\partial^P\phi\partial_P\phi\right)^{n-1}
\partial_M\phi\partial_N\phi \nonumber \\[1mm]
&& \hspace*{2cm} + 2\sum_{n=1}n\,f^{(2)}_n(\phi,\chi)\left(\partial^P\chi\partial_P\chi\right)^{n-1}\partial_M\chi\partial_N\chi
-g_{MN}\,{\cal L}_{sc}\,. \label{Tmn-2non}
\end{eqnarray}
%%%%%%%%%%%%%%
One could proceed to derive the diagonal and off-diagonal components of the
field equations and the corresponding constraints that these give, however, we
observe that, upon defining the new functions
%%%%%%%%%%%%%%%%%%%
\begin{eqnarray}
\tilde f^{(1)}(\phi,\chi) &=&
\sum_{n=1}n f_n^{(1)}(\phi,\chi) \left(\partial^M\phi\partial_M\phi\right)^{n-1}\,, \\
\tilde f^{(2)}(\phi,\chi)&=&
\sum_{n=1}n f_n^{(2)}(\phi,\chi) \left(\partial^M\chi\partial_M\chi\right)^{n-1}\,,
\end{eqnarray}
%%%%%%%%%%%%%%
the expression of the energy-momentum tensor (\ref{Tmn-2non}) reduces to the
one of Eq. (\ref{Tmn-2scalars}) with the $f^{\,(1,2)}(\phi,\chi)$ coupling functions being
replaced by $\tilde f^{\,(1,2)}(\phi,\chi)$. As the exact expressions of the coupling
functions never played a role for the existence or not of a viable solution, the
analysis presented above for the two scalar fields with canonical kinetic terms
still holds, and leads to the absence of a solution with the assumed $y$-dependence
of the mass function even in the context of the more general theory (\ref{Lagr_2non}).

%%%%%%%%%%%%%%%%%%%%%%%%%%%%%%%%%%%%%%%%%%%%%%%%%%%%%%%%%%%%%%%%%%%%

\subsection{Two interacting scalar fields with mixed kinetic terms\label{section_mixing_1}}

Let us now address a more complex model of two scalar fields $\phi$ and $\chi$ that
have minimal but mixed kinetic terms. Then, the Lagrangian reads:
%%%%%%%%%%%%%%%%%
\begin{equation}
\label{action_two_1_mixed}
{\cal L}_{sc}=f^{(1)}(\phi,\chi)\,\partial^M\phi\partial_M\phi+
f^{(2)}(\phi,\chi)\,\partial^M\chi\partial_M\chi+f^{(3)}(\phi,\chi)\,\partial^M\phi\partial_M\chi
+V(\phi,\chi)\,.
\end{equation}
%%%%%%%%%%%%%%%
For the above scalar theory, the energy-momentum tensor reads:
\begin{eqnarray} \hspace*{-0.5cm}
T_{MN}&=&2f^{(1)}(\phi,\chi)\,\partial_M\phi\partial_N\phi+2f^{(2)}(\phi,\chi)\,\partial_M\chi\partial_N\chi
\nonumber \\[1mm]
&& \hspace*{1cm} +\,f^{(3)}(\phi,\chi)\left[\partial_M\phi\partial_N\chi+\partial_M\chi\partial_N\phi\right]
-g_{MN}\,{\cal L}_{sc}\,.
\end{eqnarray}
%%%%%%%%%%%%%%%

The vanishing, off-diagonal components $(^v_{\ r}),\, (^v_{\ y})$ and~$(^y_{\ r})$ of the
field equations lead in this case to the extended constraints
\begin{eqnarray}
\label{constr_1_mixed}
&& \hspace*{-1cm}
f^{(1)}(\phi,\chi)\,(\partial_r\phi)^2+f^{(2)}(\phi,\chi)\,(\partial_r\chi)^2+f^{(3)}(\phi,\chi)\,\partial_r\phi\partial_r\chi=0\,, \\[1mm]
\label{constr_2_mixed}
&&  \hspace*{-1cm}
2f^{(1)}(\phi,\chi)\,\partial_r\phi\partial_y\phi+2f^{(2)}(\phi,\chi)\,\partial_r\chi\partial_y\chi+
f^{(3)}(\phi,\chi)\left[\partial_r\phi\partial_y\chi+\partial_y\phi\partial_r\chi\right]=0\,.
\end{eqnarray}
%%%%%%%%%%%%%%%%

As in the previous model, and following similar steps, it is easy to exclude the particular
cases where one of the fields (or both) does not depend on either $r$ or $y$.
In both cases, we are forced again to abandon the assumption of a non-trivial profile
of the mass function along the extra coordinate. Therefore, we address directly the most
general case where $\phi=\phi(v,r,y)$ and $\chi=\chi(v,r,y)$. The two off-diagonal
constraints (\ref{constr_1_mixed})-(\ref{constr_2_mixed}) can now be solved to yield
the values of two coupling functions in terms of the third one -- for example,
%%%%%%%%%%%%%%%
\begin{equation}
f^{(2)}=f^{(1)}\,\frac{(\partial_r \phi)^2}{(\partial_r \chi)^2}\,, \qquad \qquad
f^{(3)}=-2 f^{(1)}\,\frac{\partial_r \phi}{\partial_r \chi}\,.
\label{coupl_mixed}
\end{equation}
%%%%%%%%%%%%%%%%
Using the above relations, the $(yv$) and ($rv$)-components may be rewritten as
\begin{equation}
\frac{\partial_y m}{r^2}+\frac{\partial_r \partial_y m}{r}=
\frac{2f^{(1)}}{(\partial_r \chi)^2}\,(\partial_v\phi\partial_r\chi-
\partial_r\phi\partial_v\chi)\,
(\partial_y\phi\partial_r\chi-\partial_r\phi\partial_y\chi)
\label{G_yv_mixed}
\end{equation}
%%%%%%%%%%%%%
\begin{equation}
\frac{2\partial_v m}{r^2}-\frac{e^{2A}}{r} \left(4A' \partial_y m+\partial^2_y m\right)=
\frac{2f^{(1)}}{(\partial_r\chi)^2}\,(\partial_v \phi \partial_r \chi-
\partial_r \phi \partial_v \chi)^2\,,
\label{G_rv_mixed_v2}
\end{equation}
%%%%%%%%%%%%%%%
respectively.  Similarly, the rearrangement of the diagonal $(\theta\theta)$ and  $(yy)$
components, takes the form
%%%%%%%%%%%%%%
\begin{equation}
\frac{2e^{-2A}}{r^2}\,\partial_r m+ 3A'' =
-\frac{ 2 f^{(1)}}{(\partial_r\chi)^2}\,(\partial_y \phi \partial_r \chi-
\partial_r \phi \partial_y \chi)^2\,,
\label{mm-eq-new}
\end{equation}
%%%%%%%%%%%%%
We now take the square of Eq. (\ref{G_yv_mixed}) and combine it with Eqs. (\ref{G_rv_mixed_v2}) and
(\ref{mm-eq-new}). By doing so, we obtain a differential equation for the mass function with
no dependence on the fields and their coupling functions, having the form
%%%%%%%%%%%%%%%%%%
\begin{equation}
\left(\frac{\partial_y m}{r^2}+\frac{\partial_r \partial_y m}{r}\right)^2=
\left(\frac{2e^{-2A}}{r^2}\,\partial_r m+ 3A''\right)
\left[\frac{e^{2A}}{r}\left(4A' \partial_y m+\partial^2_y m\right)-\frac{2\partial_v m}{r^2}\right]\,.
\label{mass-con-mixed}
\end{equation}
%%%%%%%%%%%%
Any expression for the mass function, consistent with the complete set of equations should
satisfy the above constraint.

However, the expression of the mass function can be found through another equation
resulting from the rearrangement of the $(vv)$ and $(\theta\theta)$ components,
i.e. the following
%%%%%%%%%%%%%
\beq
 \frac{1}{r}\,\partial^2_{r}m - \frac{2}{r^2}\,\partial_{r}m =
2 f^{(1)} \partial_r\phi\,\partial_v \phi+
2f^{(2)} \partial_r \chi \partial_v \chi +
f^{(3)} (\partial_r \phi \partial_v \chi + \partial_r \chi \partial_v \phi)\,. \label{mass-eq2-mixed}
\eeq
%%%%%%%%%%%%%
Using again the relations (\ref{coupl_mixed}), one may see that the combination of terms
on the right-hand-side of the above equation vanishes, and the solution (\ref{mass-sol1})
for the mass-function is thus recovered. Substituting that solution into the constraint
(\ref{mass-con-mixed}), and equating the coefficients of the same powers of the radial
coordinate on both sides, we are led again to the constraint $\partial_y C=0$. In this case,
although a $y$-dependence remains in the expression of the mass function, this is restricted
only in the coefficient $B$ of the $r^3$ term; the Schwarzschild coefficient $C$, that is
associated to the singular terms in the curvature invariants, is not allowed to have a
non-trivial profile along the extra coordinate in contradiction with our assumptions.

We conclude this section by noting that the above analysis holds also in the case where
non-minimal forms for all kinetic terms in Eq. (\ref{action_two_1_mixed}) are included in
the theory.

%%%%%%%%%%%%%%%%%%%%%%%%%%%%%%%%%%%%%%%%%%%%%%%%%%%%%%%%%%%%%%%%%%

\section{A Field Theory of a non-minimally-coupled Scalar}

Let us finally consider the case of a non-minimally coupled scalar
field that propagates in the bulk. The action of this theory has the form
%%%%%%%%%%
\bea
\mathcal{S} = \int d^4x\,dy\,\sqrt{-g}
\left[\frac{f(\Phi)}{2\kappa^2_{5}}R-\frac{1}{2}(\nabla\Phi)^2-V(\Phi)
-\Lambda_{B}\right]\,,
\label{action}
\eea
%%%%%%%%%%
where $f(\Phi)$ is a general, smooth, positive-definite function of
the scalar field $\Phi$. The field equations following from the above
action have the covariant form
%%%%%%%%%
\bea
f(\Phi)\,\bigl(R_{MN}-\frac{1}{2}\,g_{MN}R\bigr) = \kappa^2_{5}
\bigl(\mathbb{T}^{(\Phi)}_{MN} - g_{MN}\Lambda_{B}\bigr) \,,
\label{eqs-conformal}
\eea
%%%%%%%%%
with $\mathbb{T}^{(\Phi)}_{MN}$ being the generalized energy-momentum tensor
of the scalar field defined as
%%%%%%%%%
\bea
\mathbb{T}^{(\Phi)}_{MN} \equiv \nabla_{M}\Phi\nabla_{N}\Phi -
g_{MN}\,\bigl[\frac{1}{2}(\nabla\Phi)^2+V(\Phi)\bigr] +
\frac{1}{\kappa^2_{5}}\bigl[\nabla_{M}\nabla_{N}f(\Phi) -
g_{MN}\nabla^2f(\Phi)\bigr] \,.
\label{generalized tensor}
\eea
%%%%%%%%%

As usually, we will assume that the scalar field shares the same spherical
symmetry of the spacetime but depends on all other three spacetime coordinates, i.e.
$\Phi=\Phi(v,r,y)$. Then, calculating the components of the energy-momentum
tensor (\ref{generalized tensor}) and combining them with the ones of the
Einstein tensor (\ref{einstein-tensor}), we obtain the field equations
of the theory. First, the off-diagonal components $(^v_{\ r}), (^y_{\ r}), (^y_{\ v})$
and $(^r_{\ v})$ assume the form
%%%%%%%%%%%%%
\beq
(1+f'') (\partial_{r}\Phi)^2 + f'\,\partial^{2}_{r}\Phi =0 \,,
\label{Comp-v_r}
\eeq
%%%%%%%%%%%
\vskip -3mm
%%%%%%%%%%%
\beq
(1+f'') \partial_{y}\Phi\,\partial_{r}\Phi + f'\,\partial_{y}\partial_{r}\Phi-
A' f' \partial_{r}\Phi =0\,,
\label{Comp-y_r}
\eeq
%%%%%%%%%%%
\vskip -4mm
%%%%%%%%%%%
\beq
(1+f'') \partial_{y}\Phi\,\partial_{v}\Phi + f'\,\partial_{y}\partial_{v}\Phi
- A' f' \partial_{v}\Phi - \frac{\partial_{y}m}{r}f' \partial_{r}\Phi
= \frac{f}{r}\,(\frac{\partial_{y}m}{r} +\partial_{y}\partial_{r}m)\,,
\label{Comp-y_v}
\eeq
%%%%%%%%%%%
\vskip -4mm
%%%%%%%%%%%
\bea
&& (1-\frac{2m}{r})
\bigl[(1+f'') \partial_{v}\Phi\,\partial_{r}\Phi +
f'\,\partial_{v}\partial_{r}\Phi\bigr] + (1+f'')(\partial_{v}\Phi)^2 +
f'\,\partial^{2}_{v}\Phi
- \frac{\partial_{v}m}{r}f' \partial_{r}\Phi \nonumber\\[1mm]
&& + \frac{f'}{r}\,\partial_{v}\Phi\,\bigl(\partial_{r}m-\frac{m}{r}\bigr)
+ \frac{\partial_{y}m}{r}\,e^{2A} f'\partial_{y}\Phi  =
 f \bigl[\frac{2}{r^2}\,\partial_{v}m - \frac{e^{2A}}{r}\,
(\partial^{2}_{y}m + 4A'\partial_{y}m)\bigr],
\label{Comp-r_v}
\eea
%%%%%%%%%%%
respectively. In the above, $f^{\,'}$ denotes the derivative of the coupling
function with respect to $\Phi$ and $A'$ the derivative of the warp-factor
function with respect to $y$. For simplicity, we have absorbed the
five-dimensional gravitational constant $\kappa_5^2$ inside $f(\Phi)$.
We should note here that the components $(^v_{\ y})$ and $(^r_{\ y})$ are not
independent but lead again to Eqs. (\ref{Comp-y_r}) and (\ref{Comp-y_v}),
respectively.

The diagonal components are also not all independent but lead to three
different field equations, namely
%%%%%%%%%%%%
\bea
&& \hspace*{-0.5cm} e^{-2A}\bigl[(1+f'')\,\partial_{v}\Phi\,\partial_{r}\Phi +
f'\,\partial_{v}\partial_{r}\Phi
+ \frac{f'}{r}\,\partial_{r}\Phi\,(\frac{m}{r}-\partial_{r}m)\bigr] \nonumber\\[1mm]
&&\hspace*{2cm} +A' f'\,\partial_{y}\Phi -
(\mathcal{L}_{\Phi} + \Box f+\Lambda_{B})
= f\,\bigl(6A'^2 + 3A'' - \frac{2e^{-2A}}{r^2}\,\partial_{r}m \bigr),
\label{Comp-v_v}
\eea
%%%%%%%%%%%%
\vskip -4mm
%%%%%%%%%%%%
\bea
&& \hspace*{-2cm}\frac{e^{-2A}}{r}\,f'\bigl[\partial_{v}\Phi +
\bigl(1-\frac{2m}{r}\bigr)\,\partial_{r}\Phi\bigr] +A'f'\,\partial_{y}\Phi
-(\mathcal{L}_{\Phi} +\Box f +\Lambda_{B})
\nonumber\\[1mm] && \hspace*{5cm}
=f\,\bigl(6A'^{2} + 3A^{''} - \frac{e^{-2A}}{r}\,\partial^2_{r}m\bigr),
\label{Comp-th_th}
\eea
%%%%%%%%%%%
\vskip -4mm
%%%%%%%%%%%
\bea
(1+f'')(\partial_{y}\Phi)^2 + f'\,\partial^2_{y}\Phi
- (\mathcal{L}_{\Phi} +\Box f+\Lambda_{B}) =
f\bigl[6A'^{2} - \frac{e^{-2A}}{r}\bigl(\partial^2_{r}m + \frac{2}{r}\,\partial_{r}m
\bigr)\bigr].
\label{Comp-y_y}
\eea
%%%%%%%%%%%
The quantities $\mathcal{L}_{\Phi}$ and $\Box f$ appearing above have the
explicit forms
%%%%%%%%%%%%%
\beq
\mathcal{L}_{\Phi} \equiv \frac{1}{2}(\nabla\Phi)^2+V(\Phi)=\frac{e^{-2A}}{2}\Bigl[
2\,\partial_v\Phi\,\partial_r\Phi+\bigl(1-\frac{2m}{r}\bigr)(\partial_r\Phi)^2\Bigr]
+\frac{1}{2}\,(\partial_y\Phi)^2 +V(\Phi)\,,
\eeq
and
\beq
\Box f %\equiv \frac{1}{\sqrt{-g}}\,\partial_M[\sqrt{-g} \partial^M f]
= e^{-2A}\,\partial_v\partial_rf+\frac{e^{-2A}}{r^2}\,\partial_r\Bigl[
r^2 \partial_vf+r^2\bigl(1-\frac{2m}{r}\bigr)\,\partial_rf\Bigr]
+e^{-4A}\partial_y\bigl(e^{4A}\,\partial_yf\bigr)\,,
\eeq
%%%%%%%%%%%%
respectively. An alternative constraint can be derived, that does
not contain these complicated expressions, by combining Eqs. (\ref{Comp-v_v})
and (\ref{Comp-th_th}); this has the form
%%%%%%%%%%%%
\bea
(1+f'') \partial_{v}\Phi\,\partial_{r}\Phi + f'\,\partial_{v}\partial_{r}\Phi=
\frac{f'}{r} \Bigl[\partial_{v}\Phi + (1-\frac{3m}{r}+
\partial_r m)\,\partial_{r}\Phi\Bigl] +
\frac{f}{r} \bigl(\partial^2_r m - \frac{2}{r}\,\partial_r m \bigr)\,.
\label{constraint}
\eea
%%%%%%%%%%%

One may attempt to simplify the above set of equations by assuming a simpler
configuration for the bulk scalar field $\Phi$. Unfortunately, configurations that
depend solely on one of the coordinates $(v,r,y)$ can be shown quite easily not to
be supported by the set of equations. For instance, if $\Phi$ depends only on the
radial coordinate $r$, then Eq. (\ref{Comp-y_r}) demands the triviality of either
the warp factor ($A'=0$) or the coupling function ($f^{\,'}=0$) - the former choice
removes the warping of the 5-dimensional spacetime and is thus excluded; the latter
corresponds to the minimal coupling case that was investigated in Section 4, and
shown not to lead to the desired type of solution. On the other hand, if $\Phi$
depends only on the time coordinate $v$, then Eqs. (\ref{Comp-v_v}) and
(\ref{Comp-y_y}), when combined, give the simple equation
%%%%%%%%%%%%%%
\beq
\partial^2_r m =-3 e^{2A} A'' r\,,
\eeq
%%%%%%%%%%%%%%
which upon integration yields
%%%%%%%%%%%
\beq
\partial_r m = -\frac{3}{2}\,e^{2A} A'' r^2 + m_0(v,y).
\eeq
%%%%%%%%%%%%%%%
The above when substituted in the constraint (\ref{constraint}) leads to
the relation
%%%%%%%%%%%%
\beq
\frac{\partial_v\, f}{f}- \frac{2}{r}\,m_0(v,y)=0\,.
\eeq
%%%%%%%%%%%%%
Since $f=f(v)$, the above can not be satisfied due to its explicit dependence on $r$.
Finally, if $\Phi$ depends only on the bulk coordinate $y$, the left-hand-side of
Eq. (\ref{Comp-y_v}) trivially vanishes while its right-hand-side, upon integration, yields
%%%%%%%%%%%%
\beq
m(v,r,y)=\frac{C(v,y)}{r}+D(v,r)\,.
\eeq
%%%%%%%%%%%%
The above, when substituted into Eq. (\ref{constraint}), leads to the demand that
the function $C$ be $y$-independent. However, admitting that $C$ is not a function
of the bulk coordinate contradicts our main assumption that the mass function
has a non-trivial profile along the extra dimension.
%%%%%%%%%%%%

We are thus forced to assume that the bulk scalar field depends on,
at least, a pair of coordinates. Let us investigate each case in detail:

\begin{itemize}
\item{} $\Phi$ depends on the set $(v,r)$. But then,
Eq. (\ref{Comp-y_r}) reduces again to the constraint $A' f^{\,'}=0$ -
as discussed above, none of the two choices, $A'=0$ or $f^{\,'}=0$, is
consistent with our assumptions.

\item{} $\Phi$ depends on the set $(v,y)$. Then, Eq. (\ref{constraint})
takes the simplified form
%%%%%%%%%%%%
\beq
\partial_r^2 m -\frac{2}{r}\,\partial_r m = -\frac{f'}{f}\,\partial_v \Phi\,.
\label{cons-vy}
\eeq
%%%%%%%%%%%%%
According to our assumption, the r.h.s of the above equation does not depend
on the radial coordinate - identifying it with an arbitrary function $C(v,y)$,
and integrating the above equation twice with respect to $r$, we obtain the
following expression for the mass function
%%%%%%%%%%%%
\beq
m(v,r,y)=-C(v,y)\,\frac{r^2}{2} + D(v,y)\,\frac{r^3}{3} + E(v,y)\,,
\label{mass-vy}
\eeq
%%%%%%%%%%%%
where $D(v,y)$ and $E(v,y)$ are also arbitrary functions. Also, combining
Eqs. (\ref{Comp-v_v}) and (\ref{Comp-y_y}), we obtain the constraint
%%%%%%%%%%%%
\beq
(1+f'')(\partial_{y}\Phi)^2 + f'\,\partial^2_{y}\Phi -A' f'\,\partial_{y}\Phi
+3 f A''=-\frac{f}{r}\,e^{-2A}\,\partial_r^2 m\,.
\eeq
%%%%%%%%%%%%
Since $A=A(y)$ and $f=f\bigl(\Phi(v,y)\bigr)$, the left-hand-side of the above
equation is not a function of the radial coordinate. When the expression
for the mass function (\ref{mass-vy}) found above is used, we conclude
that the only way its right-hand-side is not a function of $r$ either is to impose
the condition $C(v,y) \equiv 0$. But, from Eq. (\ref{cons-vy}), it holds
that $C(v,y) \propto \partial_v \Phi$, and according to our assumption
the latter is not zero. As a result, this case also leads to an inconsistency.

%%%%%%%%%%%%%%%%%%%%%%%%%%%%%%%%%%%%%%%%%%

\item{} $\Phi$ depends on the set $(r,y)$. In that case, Eq. (\ref{Comp-y_v})
can be rewritten in the form
%%%%%%%%%%%%%%%
\beq
(\partial_r f)\,\partial_{y}m
+ f\,(\frac{\partial_{y}m}{r} +\partial_{r}\partial_{y}m)=0\,.
\label{Comp-y_v-new}
\eeq
%%%%%%%%%%%
Similarly, the constraint (\ref{constraint}) is simplified to
%%%%%%%%%%%
\bea
(\partial _r f) \Bigl(1-\frac{3m}{r}+
\partial_r m\Bigr) +
f \bigl(\partial^2_r m - \frac{2}{r}\,\partial_r m \bigr)=0\,.
\label{constraint-new}
\eea
%%%%%%%%%%%
The above system of equations may be considered as a linear, homogeneous system for
the set of quantities $(f, \partial_r f)$: in order for it to be consistent, the determinant of the
coefficients should vanish. This leads to an equation involving only the mass function and
its derivatives, namely
%%%%%%%%%%%%%%
\beq
(\frac{\partial_{y}m}{r} +\partial_{r}\partial_{y}m) \Bigl(1-\frac{3m}{r}+ \partial_r m\Bigr)
-\partial_{y}m \, \bigl(\partial^2_r m - \frac{2}{r}\,\partial_r m \bigr)=0\,.
\label{mass-con-ry}
\eeq
%%%%%%%%%%%%%%%%%

Drawing inspiration from the form of the black-hole solutions in the context of General
Relativity or from the projected-on-the-brane analytically-known solutions, we will now
make the assumption that the mass-function can be written as the sum
of a finite number of terms each one being a power, either positive or negative, of the
radial coordinate $r$. We will thus use the general form
%%%%%%%%%%%%%
\beq
m(v,r,y)=\sum_n \, \alpha_n(v,y)\,r^n\,,
\label{general-form}
\eeq
%%%%%%%%%%%%%%%
where $n$ is an integer number taking a finite number of values. Substituting the above
expression into the constraint (\ref{mass-con-ry}) we arrive at the relation
%%%%%%%%%%%%%
\beq
\sum_\ell (\partial_y \alpha_\ell)\,r^{\ell-1} \left[(\ell+1) +
\sum_n \alpha_n\,r^{n-1} (n-\ell+1)(n-3)\right]=0\,.
\eeq
%%%%%%%%%%%%%
The above constraint is satisfied for all values of $\ell$ (and $n$) for $\partial_y \alpha_\ell=0$.
If, however, we insist that $\partial_y \alpha_\ell$ is non-trivial, then, it is the quantity inside
the square brackets that should vanish instead. The only values of $n$ for which this holds are the
$n=1$ and $n=3$. But, then, the metric function $1-2m/r=1-2\alpha_1 -2 \alpha_3 r^2$ does not
describe a black hole but a modified dS/AdS spacetime.

\end{itemize}

It is therefore the most general case, again, the one with $\Phi=\Phi(v,r,y)$, that remains
to be considered. We notice that Eq. (\ref{Comp-v_r}) can be solved to yield
%%%%%%%%%%%%%%
\begin{equation}
1+f''(\Phi)=-f'(\Phi)\,\frac{\partial_r^2\Phi}{(\partial_r\Phi)^2}\,,
\label{f''}
\end{equation}
%%%%%%%%%%%%%%
which can then be replaced into Eq. (\ref{Comp-y_r}) to obtain
%%%%%%%%%%%%%%
\begin{equation}
A'=\partial_r\left(\frac{\partial_y\Phi}{\partial_r\Phi}\right)\,.
\end{equation}
%%%%%%%%%%%%%
We may easily integrate the above differential equation with respect to $r$ to find
%%%%%%%%%%%%%%%%
\begin{equation}
\partial_y\Phi=\partial_r\Phi \bigl[A'(y)\,r+F(v,y)\bigr]\,,
\label{sol_y}
\end{equation}
%%%%%%%%%%%%%%
where $F(v,y)$ is an arbitrary function. In \cite{KPZ}, the above relation was used to
formulate a no-go theorem according to which no viable solutions with $m=m(v,y)$
exist even for an arbitrary coupling function $f(\Phi)$. However, in the present case,
the assumed $r$-dependence of the mass function impedes the formulation of a
similar argument as it makes further integrations with respect to $r$ impossible
to perform. Thus, an alternative route must be followed.

If we use Eq. (\ref{sol_y}) to substitute $\partial_y \Phi$ in Eq. (\ref{Comp-y_v}) and
then combine with the constraint (\ref{constraint}), we obtain the following differential
equation for the coupling function $f$
%%%%%%%%%%%%%%
\bea && \hspace*{-1cm}
\partial_r f \left[(A' r +F)\,\Bigl(1-\frac{3m}{r}+\partial_r m\Bigr) + r \partial_v F -
\partial_y m + F \frac{\partial_v \Phi}{\partial_r \Phi}\right] \nonumber \\[1mm]
&& \hspace*{1.5cm} +
f \left[(A'r +F)\,\Bigl(\partial_r^2 m -\frac{2}{r}\,\partial_r m\Bigr) -
\frac{1}{r}\,\partial_y m-\partial_r \partial_y m \right]=0\,.
\label{diff-f-1}
\eea
%%%%%%%%%%%%%%%%%
Also combining the diagonal components (\ref{Comp-th_th}) and (\ref{Comp-y_y}),
and using again Eq. (\ref{f''}) to substitute $(1+f'')$ in the ensuing equation, 
we obtain a second differential equation for $f$, namely
%%%%%%%%%%%%%%
%\beq
%(1+f'')\,(\partial_y \phi)^2 + f'\partial^2_y \phi - A'f'\partial_y \phi + 3f A'' =
%\frac{e^{-2A}}{r}\,f'\left[\partial_v\phi + \Bigl(1-\frac{2m}{r}\Bigr) \partial_r \phi \right]
%-\frac{2e^{-2A}}{r^2}\,f\partial_r m\,.
%\label{constraint2}
%\eeq
%%%%%%%%%%%%%%

%%%%%%%%%%%%%%
\beq
\partial_r f \left[1-\frac{2m}{r} -r e^{2A}\,(A'' r +\partial_y F)+
\frac{\partial_v \Phi}{\partial_r \Phi}\right] +
f \Bigl(-3A''r e^{2A}-\frac{2}{r}\,\partial_r m\Bigr)=0\,.
\label{diff-f-2}
\eeq
%%%%%%%%%%%%%%%%%
For the homogeneous system of Eqs. (\ref{diff-f-1}) and (\ref{diff-f-2}) to be consistent,
the determinant
of the coefficients should be zero. If we denote with $A_1$ and $A_2$ the coefficients of
$\partial_r f$ and $f$, respectively, in Eq. (\ref{diff-f-1}), and with $B_1$ and $B_2$ the
corresponding coefficients in Eq. (\ref{diff-f-2}), then we should have: $A_1 B_2 - B_1 A_2=0$.
Alternatively, we may write
%%%%%%%%%%%%%
\beq
\frac{A_1}{B_1}=\frac{A_2}{B_2}\equiv G(v,r,y)\,,
\label{ratioAB}
\eeq
%%%%%%%%%%%%%
where $G(v,r,y)$ is an arbitrary function. Focusing on the second equality of the above
relation, we may thus write
%%%%%%%%%%%%%%%
\beq
(A'r +F)\,\Bigl(\partial_r^2 m -\frac{2}{r}\,\partial_r m\Bigr) -
\frac{1}{r}\,\partial_y m-\partial_r \partial_y m =
G(v,r,y)\,\Bigl(-3A''r e^{2A}-\frac{2}{r}\,\partial_r m\Bigr)\,.
\label{ABG}
\eeq
%%%%%%%%%%%%%%

At this point, we will use again the general polynomial form (\ref{general-form})
of the mass function with respect to the radial coordinate. Employing this form,
the coefficients $A_2$ and $B_2$ also take the form of polynomials. Then, through
Eq. (\ref{ratioAB}), the arbitrary function $G$ will also be a polynomial, and thus
we may write
%%%%%%%%%%%%%%%
\beq
G(v,r,y) =\sum_\ell\,g_\ell(v,y)\,r^\ell\,,
\eeq
%%%%%%%%%%%%%%
where $\ell$ is again an integer number taking a finite number of values. Using also
the above expression in Eq. (\ref{ABG}), we obtain the constraint
%%%%%%%%%%%%%%%%
\bea && \hspace*{-2.0cm}
\sum_n \Bigl[n (n-3) \alpha_n A'-(n+1)\,\partial_y\alpha_n\Bigl] r^{n-1} +
\sum_n \alpha_n F n(n-3)\,r^{n-2} \nonumber \\[1mm]
&& \hspace*{2.5cm}=-3 \sum_\ell g_\ell A'' e^{2A} r^{\ell+1} -
2 \sum_{\ell,n} g_\ell \alpha_n n\,r^{\ell+n-2}\,.
\label{con-general}
\eea
%%%%%%%%%%%%%%%%%
The most general choice of the two integer numbers that satisfies the above equation
is $(n=3, \ell=1)$; however, in this case the metric function $1-2m(v,r,y)/r$ does not
describe any longer a black hole on the brane but a modified dS/AdS background.
More particular solutions also arise: for instance, the choice $(n=4, \ell=1)$ causes
the four terms of Eq. (\ref{con-general}) to form two pairs of the same power
of $r$ - again, the resulting metric function does not describe a black hole on
the brane. The choice $n=0$ also leads to mathematically consistent solutions,
however, from the physical point of view it fails since it leads us back to the
case with $m=m(v,y)$ studied in \cite{KPZ} or even to the black-string solutions
\cite{CHR} when the condition $\partial_y \alpha_n=0$ is imposed on top.

There are also more special cases arising from the demand that the determinant
$A_1 B_2-B_1 A_2$ trivially vanishes. This happens, for example, when the two
differential equations for $f$ reduce to one, and then $G(v,r,y)=1$. In this case,
the function $G$ does not have an $r$-dependence anymore, therefore $\ell=0$.
Equation (\ref{con-general}) still applies in this case giving either $n=2$ only or
$n=3$ only - again, in both cases, the metric function does not describe a black
hole on the brane. We should also check the case where one of the rows or columns
of the determinant has zero entries: for $A_2=B_2=0$, we obtain again the solution
(\ref{mass-sol3}) for the mass function but with $\partial_y m_0$=0 - this removes
altogether the $y$-dependence from the Schwarzschild term of the metric function
and thus the possibility of removing the bulk singularities. The case $A_1=B_1=0$,
for non-trivial $f$, unavoidably leads, through Eqs. (\ref{diff-f-1})-(\ref{diff-f-2})
to the condition $A_2=B_2=0$ and to the previous conclusion. In the case
$A_1=A_2=0$, the condition $A_2=0$ in conjunction with the general form
(\ref{general-form}) leads to the demand that $F n(n-3)=0$ -- for $F=0$, we
further obtain that $n=1$, however, either for this value or the additional
possible ones, $n=0$ and $n=3$, we are loosing again the $y$-dependence of
the metric since we are also led to the condition $\partial_y \alpha_n=0$. The
final case with $B_1=B_2=0$ is the most elaborate one: the condition $B_2=0$
leads again to the solution (\ref{mass-sol3}) for the mass function; when this
is substituted into the second condition $B_1=0$, we obtain that
%%%%%%%%%%%%%
\beq
\frac{\partial_v \Phi}{\partial_r \Phi}= \frac{2 m_0(v,y)}{r} -1 +
r \partial_y F e^{2A}\,.
\eeq
%%%%%%%%%%%%%%
Substituting both the above result and the solution (\ref{mass-sol3}) for the mass
function into the remaining equation (\ref{diff-f-1}), we obtain the constraint
%%%%%%%%%%%%%%
\bea
&& \hspace*{-1cm}
\sum_\ell f_\ell \, r^{\ell+2} \,\frac{(\ell+4)}{2}\,\partial_y \Bigl(A'' e^{2A} \Bigr) +
\sum_\ell f_\ell \,r^{\ell}\,\ell\,\Bigl(A'+\partial_v F +F \partial_y F\,e^{2A}\Bigr)
\nonumber \\[2mm]
&&\hspace*{1cm}-
\sum_\ell f_\ell\, r^{\ell-1}\,\Bigl[ 3\ell A' m_0 + (\ell+1) \partial_y m_0\Bigr] -
\sum_\ell f_\ell\, r^{\ell-2}\,\ell\,F m_0=0\,.
\label{general-con-2}
\eea
%%%%%%%%%%%%%%%
In the above, we have also assumed that, due to the polynomial form of the
coefficients $A_1$ and $A_2$, the function $f$ should also be written as a
polynomial with respect to the radial coordinate
%%%%%%%%%%
\beq
f(v,r,y) =\sum_\ell \,f_\ell (v,y) \,r^\ell\,.
\eeq
%%%%%%%%%%%%%
The mathematical consistency of Eq. (\ref{general-con-2}) demands that each one
of the terms should vanish. The most restricting is the last one leading to
$F m_0=0$: accepting that $F=0$ leads unavoidably to $A'=0$ whereas the
choice $m_0=0$ eliminates the black-hole form of the solution on the brane.

In the light of all the above, we may thus conclude that even the most general,
spherically-symmetric configuration of a non-minimally coupled bulk scalar field
cannot support a localised black-hole solution of the form (\ref{vaidya-metric}).

%%%%%%%%%%%%%%%%%%%%%%%%%%%%%%%%%%%%%%%%%%%%%%%%%%%%%%%%%%%%%%%%%%%%%%%%%%%%%%%%%%%
\section{Conclusions - Discussion}

As the quest for finding 5-dimensional, localised-on-the-brane, analytical black-hole
solutions continues, in this work we are extending a previous analysis of ours that
aimed at finding Schwarzschild-type black-hole solutions on the brane. In there
\cite{KPZ}, a generalised Vaidya-type ansatz was used for the 5-dimensional metric
tensor that allowed for both time-dependence and dependence on the bulk
coordinate of the mass function. Several types of field theories were
considered in \cite{KOT,KPZ}, that ranged from ordinary to more exotic types, in an
effort to find the field-theory framework that would support this particular type of
5-dimensional gravitational background. Unfortunately, all theories considered failed
to lead to a positive result, and that puts a big question-mark on the existence
itself of localised, close to the brane, black holes in the context of a warped
brane-world model.

The present analysis aimed at extending both analyses \cite{KOT,KPZ} by considering
again a variety of scalar field-theory models, from the simplest to the more complex
ones, but assuming a less-restrictive form for the metric tensor: instead of the
more traditional Schwarzschild form usually employed in the Vaidya metric, a more
general form was postulated for the mass function with an additional dependence on
the radial coordinate, that would allow for all possible types of modifications and
power-law terms to be present in the metric function. The increased complexity of
the metric tensor was reflected to the more elaborate form of the 5-dimensional
quantities and to the appearance of additional singular terms; however, these were
of the same type as the black-string singular terms, with their elimination being
again possible under the assumption that a solution with a fast-enough decreasing
profile of the mass function could be found.

In order to pin-point the characteristics of a field-theory model that would potentially
support such a gravitational solution, we first considered the case of a bulk filled only with
a cosmological constant. Although this model allowed for a Schwarzschild-(A)dS type
of metric function, the desired dependence of the mass-function on the bulk coordinate,
that would succeed in localising the black-hole singularities close to the brane,
was prohibited. This result holds for any sign or value of the bulk cosmological
constant. This theory had the restrictive characteristic of an isotropic energy-momentum
tensor along all five coordinates, a feature that clearly puts a restriction on the form
of the solution.

The same characteristic was present in the theory of a single scalar field with either
canonical or non-canonical kinetic term, that was studied next. Therefore, it was no
surprise when the assumed $y$-dependence of the mass function was also excluded.
Moving to the case of two interacting scalar fields, the total isotropy of the energy-momentum
tensor was at last avoided. However, the system of field equations implicitly imposed
it by demanding that the terms, that caused the differences between the components
of the energy-momentum tensor, should vanish. This result was independent of the
canonical or non-canonical type of kinetic terms of the scalar fields as well as of the
form of their common potential. The complexity of the model was increased further
by allowing a mixed kinetic term of the two fields: in this case, the isotropy of the
energy-momentum tensor was altogether avoided, and a Schwarzschild-(A)dS
type of solution was indeed found that allowed for $y$-dependence of the mass
function; unfortunately, that $y$-dependence was allowed only for the multiplicative
coefficient of the cosmological constant term and not for the one of the Schwarzschild
term associated with the bulk singular terms.

Having covered a wide variety of field theories of minimally-coupled scalar fields,
in the last part of our paper we focused on the case of a non-minimally coupled
scalar field, with a general coupling function between the scalar field and the
Ricci scalar. A bulk potential was also assumed to be present, together with the
bulk cosmological constant, however, its exact expression, once again, did not
affect the analysis in the least. Solutions for the mass-function that resembled
the ones of a Schwarzschild-(A)dS or a Reissner-Nordstrom black hole were
found again, however, the complete set of equations trivialised at the end either
the assumed $y$-dependence or the structure of the black-hole spacetime itself.

In the light of our analysis, we may conclude that allowing for a more general
form of the metric tensor, that would make space for black-hole solutions
different from the traditional Schwarzschild-type one, was not by itself capable
of producing viable solutions. The form of the field theory considered in each
case determined the form of the energy-momentum tensor, and, at times,
annihilated the increased flexibility of the field equations and took us back to
previous, singular solutions. It was only for the more involved models, of either
minimally or non-minimally coupled scalar fields, that more general black-hole
solutions were allowed, however the desired non-trivial profile of the mass-function
was excluded and thus the elimination of the bulk singularities could not take
place. Nevertheless, it became apparent that it was only in the context of specific
models, where the 5-dimensional isotropy of the energy-momentum tensor was
broken, that the $y$-dependence of the mass-function was allowed at all. As
a side observation, we should also note that the time-dependence of the scalar
field configurations seemed to be again necessary for the existence of a viable
solution, however, this feature by itself failed to produce the desired solution in
the context of the present models.

Novel solutions describing new types of black strings or regular brane-world models
have arisen in the context of our analysis, and these will be studied in detail in a
forthcoming work. However, the question of the existence of a 5-dimensional, localised
black hole in the context of a brane-world scenario still remains, and the chances of
producing one in the framework of a legitimate field theory have become even thinner.
Although valuable lessons have been learned, such as the role of the metric ansatz
and form of the energy-momentum tensor, clearly more work and inspiration are still
needed to achieve our goal.

%Based on our analysis, we have learned two important lessons: that adopting a more
%general gravitational background and a field-theory model leading to a non-isotropic
%energy-momentum tensor lead to a set of equations that are significantly less
%constrained and thus more likely to produce the desired type of black hole.

%%%%%%%%%%%%%%%%%%%%%%%%%%%%%%%%%%%%%%%%%%%%%%%%%%%%%%%%%%%%%%%%%%%%%%%%%%%%%%%%

{\bf Acknowledgments} We would like to thank Katarzyna Zuleta and Elizabeth Winstanley
for useful discussions
during the early stages of this work. This research has been co-financed by the
European Union (European Social Fund - ESF) and Greek national funds through the Operational
Program ``Education and Lifelong Learning'' of the National Strategic Reference Framework
(NSRF) - Research Funding Program: ``ARISTEIA. Investing in the society of knowledge through
the European Social Fund''. Part of this work was supported by the COST Action MP1210
``The String Theory Universe''.

%%%%%%%%%%%%%%%%%%%%%%%%%%%%%%%%%%%%%%%%%%%%%%%%%%%%%%%%%%%%%

\end{document}